# In situ digestion of canola protein gel observed by synchrotron X-Ray Scattering

Maja Napieraj[a], Annie Brûlet[a,*], Evelyne Lutton[b,c], Javier Perez[d], François Boué[a,*]

[a] *Laboratoire Léon Brillouin, UMR12 CEA-CNRS, Université Paris-Saclay, CEA Saclay, F-91191 Gif sur Yvette, France*
[b] *Mathématiques et Informatique Appliquée - Paris, UMR518 AgroParisTech-INRAE, Université Paris-Saclay, 91120 Palaiseau, France.*
[c] *Institut des Systèmes Complexes, 75013 Paris, France.*
[d] *SWING, Synchrotron SOLEIL, Saint-Aubin – BP 48, 91192 Gif sur Yvette, France.*

**Abstract**: We address the issue of structure changes of a canola protein gel (as a solid food model) during gastrointestinal digestion. We present a method for synchrotron Small-Angle X-ray Scattering analysis of the digestion of a gel in a capillary. Scanning the capillary allows tracking the digestion under diffusion of enzymatic juices. The fitting parameters characterizing the sizes, scattering intensities and structures allow to distinguish the compact, unfolded or aggregated states of proteins. The evolutions of these parameters enable to detail the complex changes of proteins during gel digestion, involving back-and-forth evolutions with proteins unfolding (1st and 3rd steps), re-compaction (2nd step) due to gastrointestinal pH and enzyme actions, before final protein scissions (4th step) resulting in small peptides. This complexity is related to the wide ranges of successive pH and enzyme activity acting on large and charged protein assemblies. Digestion is therefore impacted by the conditions of food preparation.

## 1. Introduction

Ingested protein food in human gastro-intestinal tract undergoes various physicochemical modifications, induced by pH and enzymatic environments, i.e. gastric pepsin at acidic pH~2 [1] and intestinal (mainly) trypsin at neutral pH ~7 [2], which hydrolyze peptide bonds and reduce the proteins to small peptides and free amino acids [2,3]. Structure of protein gels, resultant of the protein unfolding, aggregation and crosslinking [4], influences the enzymatic diffusion [5] and consequently digestion, in particular by slowing it down [6,7]. It is a key factor of the efficiency of digestion, thus of kinetics of nutrient absorption [8]. Although the biochemical processes of protein digestion are well described, the biophysical questions, like the conformational changes of molecular structures are little known for food made out of proteins and in particular for protein gels.

Small-Angle Scattering techniques (neutron for SANS and X-ray for SAXS) are powerful tools for analyzing soft condensed matter, and thus food [9], in a wide size range (classically from 5 Å up to 1000 Å). Digestion studies using these techniques concerned mainly lipid digestion [10,11,12] and in particular the changes of self-assembly during esterification, lipase action and hydrolysis of triglycerides.

Time resolved SAXS measurements were performed by transferring the result of in vitro digestion into a capillary [13,14]. Few studies concerned digestion of proteins. M. Bayrak et al. [15] focused on the gastric digestion of casein gels (SANS and USANS - ultra SANS), showing that the elasticity of the gel network, governed by the size and density of the protein micelles, determined the disintegration behaviour and diffusivity of pepsin. Pasquier et al. [16] studied digestion of canola proteins by SANS, showing the impact of the food structure on the digestion kinetics with differences in digestion rates between gels and protein solutions.

Here, we also study a protein gel, which is a model of solid food, made from canola proteins by Small-Angle X-ray Scattering on a synchrotron source. The technique of scanning vertically the capillary at successive times, takes advantage of the gel state, which prevents convection and enables us to follow the evolutions of the protein conformation and aggregation level.

## 2. Sample preparation & Small angle X-ray scattering

The fate of a canola protein gel (mixture of cruciferin and napin) is studied during a simulated digestion based on the standardized INFOGEST protocol [17,18]. The protein gel is formed by heating (at 95°C during 30 minutes) the protein aqueous solution (at a concentration c = 0.1g/cm$^3$ and pH 11 [19]) placed in a thin glass capillary (0.15cm in diameter).

The gastric juice solution was prepared by mixing a pepsin stock solution with water and 1 M HCl to provide a pH of 2. The intestinal solution consisted of pancreatic enzymes, bile (at 10 mM), water and sodium bicarbonate solution (2 M) providing a pH of 7. The pepsin and trypsin activities in the digestive solutions were respectively 2000 U.mL$^{-1}$ and 100 U.mL$^{-1}$ (if averaged over the digestive solution plus the whole gel volumes).

The protocol for digesting a gel in a capillary begins with a gentle injection of the gastric solution (twice the volume of the gel, i.e. 70 $\mu$L) using a long-needle syringe to ensure contact of the solution with the top of the gel without mechanically interfering in the gel itself. Capillaries were rapidly transferred to the spectrometer sample holder, then thermalized to 37±1°C to initiate digestion. Thermalization did not last more than about 20 minutes. Gastric digestion was performed first and monitored for $t_{Gas}$ of 3.5 hours. Prior to intestinal digestion, the supernatant gastric solution was rapidly removed with a syringe (leaving the volume close to the solution-gel interface, which contained digested proteins) and replaced by 70 $\mu$L of intestinal solution. SAXS measurements were then continued for long intestinal digestion $t_{Int}$ of 15.5 hours, plus an additional intestinal digestion step with freshly prepared intestinal juice, $t_{IntFresh}$ of 2 hours.

The structure of the gel is monitored by Small Angle X-ray scattering (SAXS) collected on SWING beamline at Synchrotron SOLEIL. Beam energy and sample to detector distance are set to reach a wide $q$ range from 2 10$^{-3}$ Å$^{-1}$ – 0.3 Å$^{-1}$. Measurements of the empty capillary, of a capillary with buffer solution at pH 11, of the electronic background are recorded for data treatment using the foxtrot software of the SWING beamline. Intensities were converted in absolute units (cm$^{-1}$) using a sample of water as reference.

The high flux of synchrotron allows quick (50 ms) and well spatially resolved (150 $\mu$m) vertical scanning measurements (67 steps) using a narrow beam (50 $\mu$m in height and 300 $\mu$m in width). A scan of one capillary is recorded every 35 minutes. Hence, without precedent, a huge amount of data was collected as a function of time $t$ and capillary position $z$, during *in situ* gastrointestinal digestion of about 20 hours. The effect of radiation damage was checked. It was negligible compared with the observed variations in the gel structure.

## 3. Results and discussion

An example of kinetics of the small-angle X-ray scattering at one $z$ position of the capillary during about 20 h of digestion is shown in **Fig. 1**, on a log-log scale to emphasize the evolution of the gel structure. What appears first, is the gradual decay of the scattering intensity at low $q$ and a non-monotonous variation at higher $q$. In order to quantify the evolution of the gel structure, all spectra are fitted with a "Two-Lorentzian" function [20] described by

$$I(q) = \frac{A}{1+(q\xi)^n} + \frac{C}{1+(q\Xi)^m} + Bkg \qquad \text{Eq. 1}$$

where the 1$^{st}$ term corresponds to the high $q$ range, which should figure out the scattering from the proteins-size objects. The 2$^{nd}$ term corresponds to low the $q$ range and should characterize the scattering from aggregated proteins. $\xi$ and $\Xi$ are the correlation lengths of the gel components (proteins, aggregates) and $n$ and $m$ exponents indicating their compaction states. Parameters $A$ and $C$ are the respective intensity factors and $Bkg$ is a constant (weak to account for a remaining background signal).

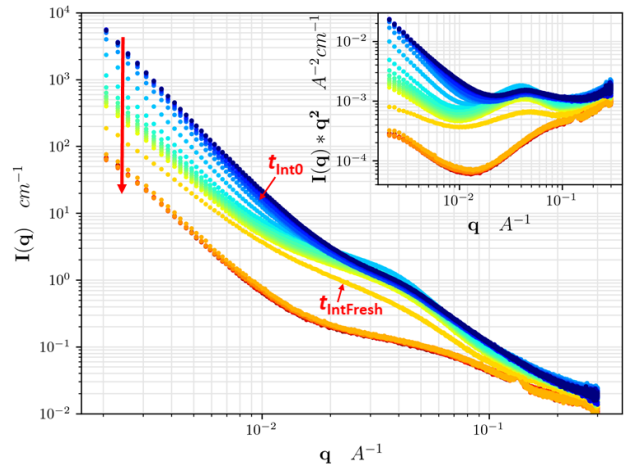

**Fig. 1.** SAXS spectra presenting the digestion kinetics of canola protein gel for one z position in the capillary. Spectra have been recorded with 35minutes. Colors denote the digestion time and red arrows are showing its direction; first 4 curves present 2.2 hours of gastric digestion step, 5$^{th}$ curve the injection of intestinal juice, and the following curves are for 15.5 hours of intestinal digestion, plus 2 hours with fresh intestinal enzymes. At long intestinal digestion times, a small peak at 0.133 Å$^{-1}$ appears: it can be related to some conjugation of bile salts present in the intestinal juice with digestion products (like amino-acids, for example glycine). Inset: Kratky representations of the same data.

### 3.1 Digestion at the protein scale: a ($\xi$, n) diagram

In this paragraph, we present the changes in the fitting parameters corresponding to the size range of proteins: the correlation length $\xi$, the exponent $n$, and the "intensity" ($A$ in Eq. 1). Through the parameters for several ($z$, $t$) pairs, we reconstruct a complete protein digestion process, progressing in the gel with time in a diffusion-reaction manner. The same protein evolutions appear delayed in time for $z$ positions further from the gel /digestive juice interface. The results of this analysis in the ($\xi$, $n$, **intensity**) space gather on master curves shown in **Fig. 2**. These bring the validity of our analysis to a higher level.

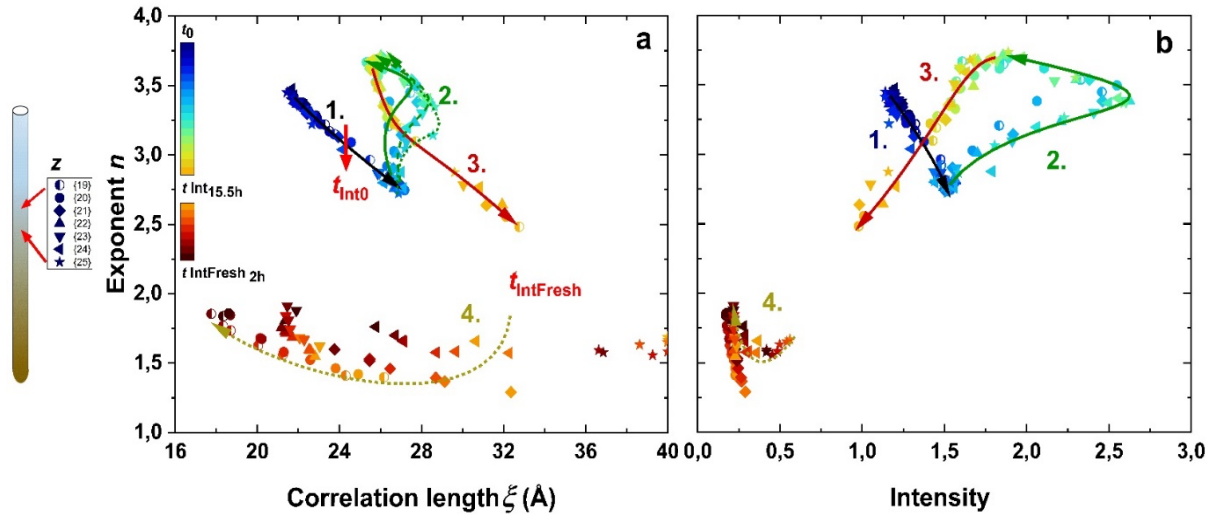

**Fig. 2.** Diagrams presenting the evolution during the digestion of the fit parameters ($\xi$, **n**, intensity) describing the structural changes at the scale of proteins. **a)** Exponents **n** as a function of correlation length $\xi$. **b)** Exponent **n** as a function of intensity factor. Symbols shapes correspond to seven different z positions in the capillary (z =19-25, with a 150 µm interval) and colors correspond to the digestion time as in **Fig. 1**. Numbers indicate the four characteristic steps of digestion and arrows, the evolution directions.

Four steps describing the complex process of digestion of protein at local scale are observed.

### 1$^{st}$ step: Protein unfolding/ Aggregation

Scattering from the proteins in the gel before digestion (1$^{st}$ navy blue curve in **Fig. 1**) shows rather folded conformations revealed by an oscillation for $q > 0.02$ Å$^{-1}$, due to the toroidal shape of cruciferin (predominating in the scattering signal due due its high mass compared to the one of napin). We find $\xi = 22$ Å, close to $R_{g0}/3^{1/2}$, $R_{g0} = 39$ Å, being the radius of gyration of native cruciferin (data not shown).

Let us make a simplistic analogy with a polymer solution. With a mass of cruciferin $M_0 = 300$ kDa and a native $R_{g0} = 39$ Å, the overlapping concentration between two proteins $c^* = M_0/N_{av} / \left(\frac{4\pi}{3} R_{g0}^3\right)$ is about 2g/cm$^3$. In the gel at 0.1g/cm$^3$, cruciferin is well below c*. Even if proteins are connected to each other in the gel, we use a solution-like description: the elementary meshes of the gel scatter as non-overlapping entities containing one protein, like end-linked chains in a gel following the c* theorem [21]. **n** ~3.4, close to 4, indicates a compact structure with sharp interfaces (Porod law [22]), here the folded protein. During the 3.5 h of gastric digestion at very acidic pH, we record subtle changes: a decrease of **n** from ~3.4 to 3.2, and a slight increase of $\xi$ from ~22 Å to 24 Å, indicating a beginning of unfolding of proteins. That is faster after injection of intestinal juice at pH 7 (red arrow $t_{Int0}$ in **Fig. 2a** at $t_{Gas}=$ 3.5 h): $\xi$ reaches ~27 Å and **n** ~2.8 in the first 35 min of intestinal digestion ($t_{Int0.5h}$). This indicates an ongoing unfolding of the proteins in the intestinal conditions. Another precious indication is the slight increase of intensity (see **Fig. 2b**). It cannot be due to unfolding of one protein alone in the mesh: that, on the contrary, would decrease intensity via hydration. Therefore, the observed features rather correspond to an aggregation, which we call local (i.e., involving a few proteins or protein units), arising from partially unfolded proteins. Such local aggregation is favored when the pH changes: (i) from initial 11 to gastric 2, where both component proteins (cruciferin and napin) likely cross their isoelectric points [19], favoring their unfolding and further interactions; (ii) from 2 to the intestinal pH 7, when the two proteins can attract each other due to their opposite charges. In short, for initially rather compact proteins, unfolding dominates the first step.

### 2$^{nd}$ step: Re-compaction/Aggregation

Shortly after the intestinal juice was added, the exponent **n**, surprisingly, turns back to an increase, reaching a maximum value of ~ 3.8 at $t_{Int4h}$, higher than before digestion, indicating more compact structures. Detailed changes in compactness are magnified through variations in the $q^2 I(q)$ representation at large $q$. On the inset of **Fig. 1** , we can see a sharper maximum (on the lighter blue curve), corresponding to an increased compaction of the proteins. During this entire step, the fit parameters ($\xi$, **n**, intensity) allow us to assume a non-overlapping proteins regime. In a first part, the intensity increases noticeably, up to 2.5 time the initial value. The $\xi$ variation depends on the z position: for the upper z values, $\xi$ value remains constant, while for deeper ones, it increases. These intensity and $\xi$ variations combined with ***n*** approaching 4 can be interpreted as a growth of the local aggregates together with a compaction at the scale of proteins. In a second part, while ***n*** continues increasing, both intensity and $\xi$ decrease: this can correspond to disaggregation and/or protein chain scission upon enzymatic action. Looking with more details at the first part, we observe that for deeper *z* positions, the paths in the ($\xi$, ***n***) diagram are more sinuous (see **Fig. 2a**, the 3 green curves). The explanation for this sinuosity can be that, for the deeper *z* positions, protein unfolding has continued for longer before the enzymatic scission, giving a possibility for interactions with hydrolyzed peptides, which have diffused from the above *z* positions upon digestion. This may result in formation of local assemblies, which slightly increase in size $\xi$ before the enzymatic attack and their apparent re-compaction. Furthermore, the enzymatic diffusion through such altered local structure and their subsequent action can be, in this case, slowed down. Together with a decrease of enzymatic activity with time, it can lower digestion efficiency. At the end of this re-compaction step, $\xi$, ***n***, intensity values for all *z* are similar: the final compaction state is independent on digestion kinetics in this step.

### 3$^{rd}$ step: Unfolding/Disaggregation

After reaching the maximum value ~3.8 at $t_{Int4h}$, the exponent ***n*** starts to decrease and $\xi$ is back at increasing again, making by that a sort of loop in the ($\xi$, ***n***) diagram. The ongoing evolution of the local structures comes again to the unfolding of proteins. However, this time, the intensity is getting gradually reduced, which altogether suggests a disaggregation on the local scale, through enzymatic hydrolysis, and transformation into more extended structures. This process extends, slower as time passes, up to 15.5 h of intestinal digestion. Such slowing down of protein unfolding is however not intuitive: we would rather expect an acceleration of digestion, since the access of enzymes to substrates should be favoured by the unfolding. The plausible explanation for this observation can be a lower enzymatic activity at these long intestinal digestion times.

At the end of this step, the local structure is described by the highest $\xi$ of 33 Å, ***n*** of 2.5 and intensity back the value before digestion. $\xi$ approaches the size of very unfolded proteins obtained after chemical denaturation (not shown), assuming $R_g^2 = 3.\xi^2$. The overlapping concentration would be 0.64g/cm$^3$, such that unfolded proteins are still under *c\**. Thus, the decrease of intensity is not due to interpenetration but to a lower protein mass in the mesh. Besides, the exponent ***n*** of 2.5 indicates that the correlations at intermediate scale are still present, suggesting a kind of branched polymers. At the end of this step, we therefore assume that the digestion residues are ramified clusters of sequences of proteins of rather extended conformations, polydisperse in mass.

### 4$^{th}$ step: Scission when adding fresh enzymes

At time $t_{Int15.5h}$ of intestinal digestion, the addition of fresh enzymes rapidly leads to disappearance of the cruciferin form factor oscillation in the $q^2I(q)$ plots (see **Fig. 1 inset**). The ***n*** values obtained are now as low as 1.8 (see **Fig. 2**), describing unfolded structures resembling polymer chains in a good solvent [23] (similar to what was obtained after chemical denaturation – not shown). The intensity is here strongly decreased (by a factor in between 2 and 4), and from now on, $\xi$ is also decreasing from 32Å to 18Å. Both polymer-like conformation and small size could be the results of an advanced digestion. However, since $\xi$ is very small, the low intensity could also be attributed to a strong interpenetration of protein chains, assuming different proteins to form long chains. This, however, seems unlikely. More realistic is the picture where proteins are cut into smaller parts. Such renewed digestion efficiency is probably driven by the higher activity of fresh enzymes, enhanced by a better access of enzymes to peptide bonds, once proteins are unfolded.

The final value of $\xi$ around 18 Å corresponds to the size of small peptides (with ~5 amino acids, assuming 3.5 Å per one). This agrees with a peptidomic study showing that most peptides in human jejunum (middle intestinal part) have indeed between 2 and 14 amino acids [24].

To briefly summarize, the evolution of local conformation of proteins during gastric then intestinal digestion is schematically represented in **Fig. 3**. Digestion processes involve back-and-forth evolutions with proteins unfolding (1$^{st}$ and 3$^{rd}$ steps) and re-compaction (2$^{nd}$ step) due to gastrointestinal pH conditions and enzymatic actions, before final protein scissions (4$^{th}$ step) resulting in small peptides.

### 3.2 Digestion at the large aggregates scale

Alongside those local conformational changes, we can look at the aggregates size range (above 100 nm). This corresponds to the low *q* scattering, for which we observe a continuous decrease in intensity (by factor ~ 60 at lowest *q*) and in exponent (from 3.8 down to 3). The initial values of exponent ***m*** close to 4, with a strong visible upturn towards lower *q*, seen for the undigested gel, characterize, in overall, strong aggregation. It can result from (i) a dense packing of the aggregates with well-defined interfaces, or from (ii) strong sample heterogeneity with regions poor and rich in proteins.

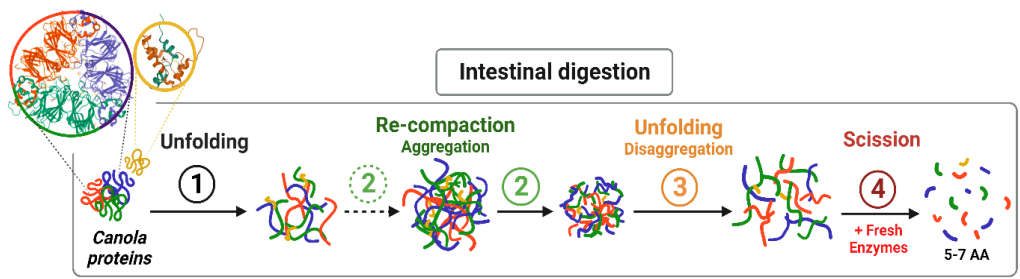

**Fig. 3.** Schematic representations of the evolution during gastrointestinal digestion of canola proteins conformation at local scale in a gel prepared at pH 11. Cruciferin is shown by the three colored proteins and napin is shown in yellow.

Under digestion, the decrease in intensity could be a sign of an enzymatic destruction of some large aggregates, leading to their disconnection, and hence destruction of the gel network. More precisely, in the frame (i) (of compact objects with sharp interfaces), the decrease with time of the scattering intensity can be attributed to a decrease in the aggregates specific area or/and in their number fraction. To make the distinction, we need to determine the aggregate sizes, which is unfortunately not accessible within this $q$ range. The progressive change of the gel scattering intensity, closer to a $q^{-3}$ decay under intestinal digestion, reflects loosening of the internal structure of the aggregates, i.e., objects of lower fractal dimension with increased interfacial roughness.

The general degradation of the large aggregates can be consistent with the intensity increase at high $q$, observed during the unfolding and re-compaction steps, if reflecting a situation where the digested proteins, released from the aggregates, re-assemble together into new, more or less compact structures.

## 4. Conclusion

Studying by synchrotron SAXS the digestion of the protein gel inside a capillary, i.e., with only a 1D enzymatic diffusion without flow or agitation, allows us to detail the main steps of the protein conformation evolution throughout almost 20 hours of digestion. Even though those evolutions are very slow, we believe that the different processes described here are meaningful. Indeed, firstly, we have verified that the enzymes remained active (although much less) up to 20 hours. Furthermore, in comparison, our previous experiments by Small Angle Neutron Scattering on the same gel type but digested in more "real-life" conditions (millimetric gel pieces, 3D enzymatic diffusion), gave a stronger gastric digestion effect already after 30 min, and even enhanced for the intestinal digestion [19]. Although the digestion conditions and kinetics were different, the local pH in the capillary (in spite of protons fast diffusion) being possibly slowly buffered due to protein rearrangements, we find undubious similarities in the protein evolutions: gastric unfolding and intestinal re-compaction, as well as in the back-and-forth variations in the ($\xi$, $n$) diagram. Moreover, particularly throughout the intestinal phase, much faster and more effective than the gastric one, SAXS unveils the complexity of the digestion processes, unattainable with our previous investigations. It is also worth to bring up that for the chosen gel, composed of initially compact proteins, gastric unfolding seems to be the first, necessary step prior to the enzymatic action.

Additional measurements show that changing the initial conformation of proteins (by gelation at different pH) influences the process and kinetics of protein digestion. Hence, the initial state of food (here proteins) could be adapted to the variations of pHs and enzyme activities, large and characteristics of each individual, including people suffering from digestive diseases. We thereby believe that the use of synchrotron SAXS, with a very detailed monitoring of the evolutions of protein sizes and shapes, may give key features for understanding digestion of many other different protein-based solid food.

**Acknowledgements**

We are grateful for the Synchrotron radiation facility SOLEIL (Beamline SWING). We thank the GDR SLAMM CNRS INRAE for sharing a Blocked Allocation Group on SWING.
This work benefited from the use of the SasView application, originally developed under NSF award DMR-0520547. SasView contains code developed with funding from the European Union's Horizon 2020 research and innovation programme under the SINE2020 project, grant agreement No 654000.